\documentclass[aps, prd, preprint, groupedaddress, nofootinbib]{revtex4-1}
\usepackage{amsmath, bm, bbold, dsfont, rotating, hyperref, footmisc, array}
\usepackage{caption,subcaption}
\usepackage{multirow}
\usepackage{tikz,pgf,pgfplots}
\usetikzlibrary{patterns}

\newcolumntype{C}[1]{>{\centering\let\newline\\\arraybackslash\hspace{0pt}}p{#1}}
\newcommand{\abs}[1]{\lvert #1 \rvert}

\begin{document}
\title{Initial Systematic Investigations of the Landscape of Low Layer NAHE Variation Extensions}
\author{Timothy Renner}
\email{Renner.Timothy@gmail.com}
\author{Jared Greenwald}
\email{Jared\_Greenwald@baylor.edu}
\author{Douglas Moore}
\email{Douglas\_Moore1@baylor.edu}
\author{Gerald Cleaver}
\email{Gerald\_Cleaver@baylor.edu}
\affiliation{Baylor University}
\date{\today}

\preprint{BU-HEPP-11-07}

\begin{abstract}
The discovery that the number of physically consistent string vacua is on the
order of $10^{500}$ has prompted several statistical studies of string
phenomenology. Focusing on the Weakly Coupled Free Fermionic String (WCFFHS)
formalism, we present systematic extensions of a variation on the NAHE
(Nanopoulos, Antoniadis, Hagelin, Ellis) set of basis vectors. This variation is
more conducive to the production of ``mirrored" models, in which the observable
and hidden sector gauge groups (and possibly matter content) are identical. This
study is parallel to the extensions of the NAHE set itself \cite{Renner:2011},
and presents statistics related to similar model properties. Statistical
coupling between specific gauge groups and spacetime supersymmetry (ST SUSY)
is also examined. Finally, a model with completely mirrored gauge groups is
discussed.
\end{abstract}

\maketitle

\section{Introduction}\label{section:introduction}
The large number of string vacua  \cite{Bousso:2000, Ashok:2003} has prompted
both computational and analytical examinations of the landscape, e.g.
\cite{Dijkstra:2004, Donagi:2004, Valandro:2008, Balasubramanian:2008,
Lebedev:2008, Gmeiner:2008, Dienes:2008, Gabella:2008, Donagi:2008}. The Weakly
Coupled Free Fermionic Heterotic String (WCFFHS) \cite{Antoniadis:1986,
Antoniadis:1987, Kawai:1986_2,Kawai:1988} approach to string model construction
has produced some of the most phenomenologically realistic string models to date
\cite{Cleaver:1999, Lopez:1992, Faraggi:1989, Faraggi:1992, Antoniadis:1990,
Leontaris:1999, Faraggi:1991, Faraggi:1992_2, Faraggi:1992_3, Faraggi:1991_2,
Faraggi:1991_3, Faraggi:1995, Faraggi:1996, Cleaver:1997, Cleaver:1997_2,
Cleaver:1997_3, Cleaver:1998, Cleaver:1998_2, Cleaver:1998_3, Cleaver:1998_4,
Cleaver:1998_5, Cleaver:1999_2, Cleaver:1999_3, Cleaver:1999_4, Cleaver:2000,
Cleaver:2000_2, Cleaver:2001, Cleaver:2001_2, Cleaver:2002, Cleaver:2002_2,
Cleaver:2002_3, Perkins:2003, Perkins:2005, Cleaver:2008, Greenwald:2009,
Faraggi:2010a, Gato-Rivera:2010a, Gato-Rivera:2010b, Gato-Rivera:2010c,
Gato-Rivera:2011a, Faraggi:2011a, Cleaver:2011}. The present study focuses on
systematic extensions of a NAHE Variation, first presented in
\cite{Greenwald:2009} and is a parallel study to the work presented in
\cite{Renner:2011}.

\subsection{The NAHE Variation}\label{section:nahe_variation}
While there have been many quasi-realistic models constructed from the NAHE
basis, other bases can be used to create different classes of realistic and
quasi-realistic heterotic string models.  Like the NAHE set, the NAHE variation
is a collection of five order-2 basis vectors. However, the sets of matching
boundary conditions are larger than those of the NAHE set. This allows for a new
class of models with ``mirrored" groups - that is, with gauge groups that occur
in even factors. Some also have mirrored matter representations that do not
interact with one another. This means that hidden sector content matches the
observable sector, making the dark matter and observable matter gauge charges
identical. Several scenarios with mirrored dark matter have been presented as
viable phenomenological descriptions of the universe \cite{Mohapatra:1996,
Mohapatra:1999, Mohapatra:2000, Mohapatra:2001}.

The NAHE set does not have a tendency to produce mirrored models because the
boundary conditions making up the $SU(4)^3$ gauge groups break the mirroring
between the elements $\overline{\psi}, \overline{\eta}$ and $\overline{\phi}$.
We can remedy this by ensuring that the worldsheet fermions
$\overline{\psi}^{1,\dots,5}$ and $\overline{w}^{1,\dots,6}$ have the same
boundary conditions as $\overline{\phi}^{1,\dots,8}$. In doing so, the NAHE
variation basis vectors generate a model with gauge group $SO(22)\otimes
E_6\otimes U(1)^5$. The basis vectors making up this set are presented in
\autoref{table:nahe_variation_basis} with the resulting particle content of the
NAHE variation model presented in \autoref{table:nahe_variation_matter}.

\begin{table}[ht]
  \caption{The basis vectors and GSO coefficients of the NAHE variation arranged
  into sets of matching boundary conditions. The worldsheet fermions $\psi$,
  $x^i$, $\overline{\psi}^i$, $\overline{\eta}^i$, and $\overline{\phi}^i$ are
  expressed in a complex basis, while $y^i$, $w^i$, $\overline{y}^i$, and
  $\overline{w}^i$ are expressed in a real basis.}

  \begin{center}
    \begin{tabular}{||c|c||c|c|c|c|c|c|c|c|c||c|c|c|c||}
      \hline\hline
      Sec & O & $\psi $ & $x^{12}$ & $x^{34}$ & $x^{56}$ & $\overline{\psi}^{1,...,5}$ & $\overline{\eta}^1$ & $\overline{\eta}^2$ & $\overline{\eta}^3$&$\overline{\phi}^{1,...,8}$&$y^{~12}||\overline{y}^{~12}$&$y^{~34}||\overline{y}^{~34}$&$y^{~56}||\overline{y}^{~56}$&$w^{1,...,6}||\overline{w}^{1,...,6}$\\
      \hline\hline
      $\vec{\mathds{1}}$&2&1&1&1&1&1,...,1&1&1&1&1,...,1&1$||$1&1$||$1&1$||$1&1,...,1$||$1,...,1\\
      \hline
      $\vec{S}$&2&1&1&1&1&0,...,0&0&0&0&0,...,0&0$||$0&0$||$0&0$||$0&0,...,0$||$0,...,0\\
      \hline
      $\vec{b}_1$&2&1&1&0&0&1,...,1&1&0&0&0,...,0&0$||$0&1$||$1&1$||$1&0,...,0$||$0,...,0\\
      \hline
      $\vec{b}_2$&2&1&0&1&0&1,...,1&0&1&0&0,...,0&1$||$1&0$||$0&1$||$1&0,...,0$||$0,...,0\\
      \hline
      $\vec{b}_3$&2&1&0&0&1&1,...,1&0&0&1&0,...,0&1$||$1&1$||$1&0$||$0&0,...,0$||$0,...,0\\
      \hline \hline
    \end{tabular}
    \vspace{5mm}\\
    \setlength{\tabcolsep}{6pt}
    $k_{ij}$ = $\left(
      \begin{tabular}{c|ccccc}
        &$\vec{\mathds{1}}$&$\vec{S}$&$\vec{b}_1$&$\vec{b}_2$&$\vec{b}_3$\\
        \hline
        $\vec{\mathds{1}}$  & 1 & 0 & 1 & 1 & 1 \\
        $\vec{S}$           & 0 & 0 & 0 & 0 & 0 \\
        $\vec{b}_1$         & 1 & 1 & 1 & 1 & 1 \\
        $\vec{b}_2$         & 1 & 1 & 1 & 1 & 1 \\
        $\vec{b}_3$         & 1 & 1 & 1 & 1 & 1 \\
      \end{tabular}
    \right)$
  \end{center}
  \label{table:nahe_variation_basis}
\end{table}

\begin{table}[ht]
  \caption{The particle content for the NAHE variation model. The model also has
  five $U(1)$ groups and $N=1$ ST SUSY.}
  \begin{center}
    \begin{tabular}{||C{1.5cm}|C{1.5cm}|C{1.5cm}||}
      \hline\hline
      \textbf{QTY}&$SO(22)$&$E_6$\\
      \hline\hline
      30 & $22$ &$1$\\
      \hline
      15 & $1$  &$27$\\
      \hline
      90 & $1$  &$1$\\
      \hline
      15 & $1$  &$\overline{27}$\\
      \hline \hline
    \end{tabular}
    \label{table:nahe_variation_matter}
  \end{center}
\end{table}

The observable sector is generally regarded as being the $E_6$, however,
contributions to the observable sector may come from the breaking of the
$SO(22)$. As compared to the NAHE set, the large number of $U(1)$s and
non-Abelian singlets is less phenomenologically favorable; however, the
quantities of both can reduce drastically which is shown in the statistics for
single layer extensions.

In \autoref{section:order_2}, layer $1$, order $2$ ($L1O2$) extensions of the
NAHE variation are investigated, with a focus on statistics. In
\autoref{section:order_3}, $L1O3$ extensions are similarly examined. In
\autoref{section:gut_models}, the statistics of GUT and of spacetime
supersymmetries of both orders are determined. \autoref{section:mirroring}
offers an example of a near mirrored model and \autoref{section:conclusion}
reviews the findings of the prior sections.

\section{Layer 1, Order 2 Extensions}\label{section:order_2}
There were $309$ quasi-unique models out of $1,315,328$ total consistent models
built given the input parameters. A redundancy related to the rotation of the
gauge groups, discussed in detail in \cite{Renner:2011}, is also present.
Duplicate models within the set of $309$ were removed by hand. Approximately
$2\%$ of the models in the data set without rank cuts were duplicates, while
none of the models with rank cuts had duplicates. The gauge group content of
those models are presented in \autoref{table:nahe_variations_l1o2_groups}.

The most common gauge group in this data set is $U(1)$, while the most common
non-Abelian gauge group is $SU(2)$, though less than half of the models contain
it. The other pertinent feature of these models is the presence of non-simply
laced gauge groups with high rank. The $SO(2n+1)$ groups range from rank $2$ up
to rank $10$. Finally, about one third of the models retain their $E_6$
symmetry. The stability of the $E_6$ is in contrast to the more common breaking
of $SO(10)$, the observable sector, in NAHE-based models \cite{Renner:2011}.
These models will be revisited later with the $E_6$ treated as an observable
sector gauge group, and the number of chiral matter generations they have will
be statistically examined.

Also of interest regarding the gauge group content of this data set is the
number of gauge group factors present in each model, see
\autoref{figure:nahe_variations_group_factors}. The distribution of the number
of gauge group factors across the unique models peaks around $8$, suggesting
that, roughly, the most common effect of $L1O2$ extension is the breaking of
only one group factor. In a few models, some of the factors have enhancements,
typically the $U(1)$ groups. Additional adjoint content distributions are
provided in \autoref{figure:nahe_variations_u1_factors}, with GUT model
distributions presented in \autoref{table:nahe_variations_gut_groups}, but will
not be discussed in detail here.

Regarding the matter content, the number of ST SUSYs is plotted in
\autoref{figure:nahe_variations_st_susys}, and the number of non-Abelian
singlets is plotted in \autoref{figure:nahe_variations_na_singlets_l1o2}. It is
clear from the latter that the number of non-Abelian singlets can get quite
high. While most models have between $50$ and $80$, there can be up to $250$
non-Abelian singlets in a model. This implies that many models in this data set
cannot be viable candidates for quasi-realistic or realistic models.

\section{Layer 1, Order 3 Extensions}\label{section:order_3}
As was the case with the NAHE extensions, there are more distinct NAHE variation
$L1O3$ extensions than $L1O2$ extensions. Out of $442,272$ models built
$1,166$ of them were unique. Based on the order-2 redundancies, the systematic
uncertainty for this data set is estimated to be $2\%$. Their gauge group
content is tabulated in \autoref{table:nahe_variations_l1o3_groups}.

As was the case with the $L1O2$ data set, $U(1)$ is the most common gauge group.
However, the percentage is significantly lower here, about $86\%$ as opposed to
$98\%$. This suggests that some of the added basis vectors are unifying the five
$U(1)$s in the NAHE variation into larger gauge groups. Also of note is the
number of models with gauge groups of rank higher than $11$. In the $L1O2$ data
set, there were only three models of this type, about $1\%$. In the $L1O3$ data
set, there were $28$ models with this property, about $2.4\%$.

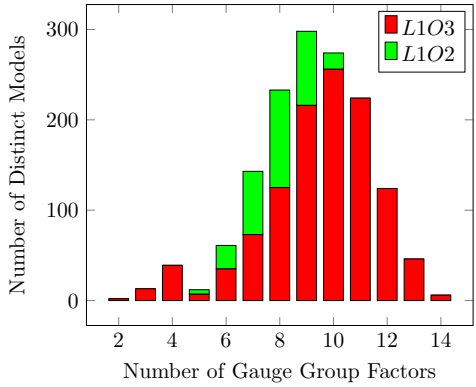
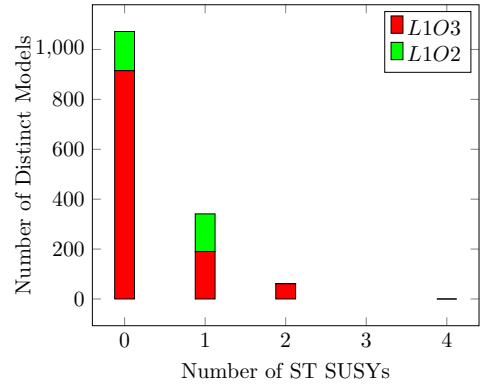
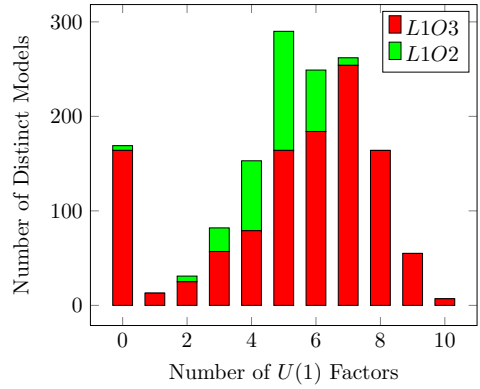
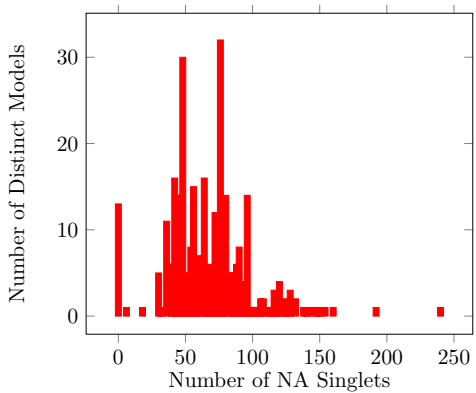
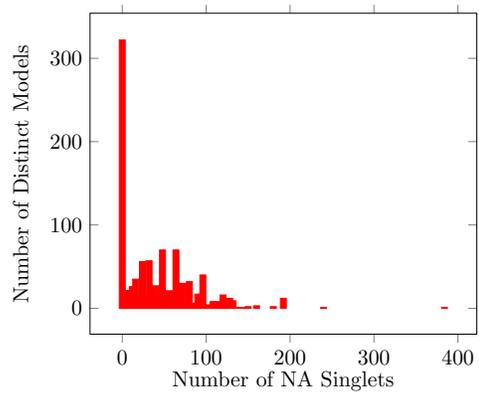
\begin{figure}[!ht]
  \centering
    \begin{subfigure}[b]{0.4\textwidth}
      \begin{tikzpicture}[scale=0.75]
        \begin{axis} [ybar stacked, ylabel=Number of Distinct Models,
                      xlabel=Number of Gauge Group Factors]
          \addplot[draw=black, fill = red, bar shift=-0pt]coordinates{
            (2,2) (3,13) (4,39) (5,7) (6,35) (7,73) (8,125) (9,216) (10,256)
            (11,224) (12,124) (13,46) (14,6)
          };
          \addlegendentry{$L1O3$}
          \addplot[draw=black, fill=green, bar shift=0pt]coordinates{
            (2,0) (3,0) (4,0) (5,5) (6,26) (7,70) (8,108) (9,82) (10,18)
            (11,0) (12,0) (13,0) (14,0)
          };
          \addlegendentry{$L1O2$}
        \end{axis}
      \end{tikzpicture}
      \caption{Number of Gauge Group Factors}
      \label{figure:nahe_variations_group_factors}
    \end{subfigure}
    \hspace{0.1\textwidth}
    \begin{subfigure}[b]{0.4\textwidth}
      \begin{tikzpicture}[scale=0.75]
        \begin{axis} [ybar stacked, ylabel=Number of Distinct Models,
                      xlabel=Number of ST SUSYs]
          \addplot[draw=black, fill=red]coordinates{
            (0,915) (1,190) (2,61) (4,0)
          };
          \addlegendentry{$L1O3$}
          \addplot[draw=black, fill=green]coordinates{
            (0,157) (1,151) (2,1) (4,0)
          };
          \addlegendentry{$L1O2$}
        \end{axis}
      \end{tikzpicture}
      \caption{Number of Spacetime SUSYs}
      \label{figure:nahe_variations_st_susys}
    \end{subfigure}
    \vspace{5mm}\\
    \begin{subfigure}[b]{0.4\textwidth}
      \begin{tikzpicture}[scale=0.75]
        \begin{axis} [ybar stacked, ylabel=Number of Distinct Models,
                      xlabel=Number of $U(1)$ Factors]
          \addplot[draw=black, fill=red, bar shift=0pt]coordinates{
            (0,164) (1,13) (2,25) (3,57) (4,79) (5,164) (6,184) (7,254) (8,164)
            (9,55) (10,7)
          };
          \addlegendentry{$L1O3$}
          \addplot[draw=black, fill=green, bar shift=0pt]coordinates{
            (0,5) (1,0) (2,6) (3,25) (4,74) (5,126) (6,65) (7,8) (8,0)
            (9,0) (10,0)
          };
          \addlegendentry{$L1O2$}
        \end{axis}
      \end{tikzpicture}
      \caption{Number of $U(1)$ Factors}
      \label{figure:nahe_variations_u1_factors}
    \end{subfigure}
    \vspace{5mm}\\
    \begin{subfigure}[b]{0.4\textwidth}
      \begin{tikzpicture}[scale=0.75]
        \begin{axis} [ybar, ylabel=Number of Distinct Models,
                      xlabel=Number of NA Singlets, bar width = 3pt]
          \addplot[draw=red, fill=red]coordinates{
            (0,13)(6,1)(18,1)(30,5)(32,1)(36,11)(40,6)(42,16)(44,14)(46,4)
            (48,30)(50,5)(52,1)(54,8)(56,15)(58,2)(60,7)(64,16)(66,2)(68,6)
            (70,3)(72,12)(74,1)(76,32)(78,4)(80,14)(84,5)(86,5)(88,6)(90,8)
            (92,4)(94,2)(96,14)(98,1)(100,1)(102,1)(104,1)(106,2)(108,2)(112,1)
            (114,1)(116,3)(118,1)(120,4)(124,2)(126,1)(128,3)(132,2)(138,1)
            (140,1)(144,1)(148,1)(150,1)(154,1)(160,1)(192,1)(240,1)
          };
        \end{axis}
      \end{tikzpicture}
      \caption{Number of Non-Abelian Singlets - $L1O2$ Extension}
      \label{figure:nahe_variations_na_singlets_l1o2}
    \end{subfigure}
    \hspace{0.1\textwidth}
    \begin{subfigure}[b]{0.4\textwidth}
      \begin{tikzpicture}[scale=0.75]
        \begin{axis} [ybar, ylabel=Number of Distinct Models,
                      xlabel=Number of NA Singlets, bar width = 3pt]
          \addplot[draw=red, fill=red]coordinates{
            (0,322)(2,7)(4,4)(6,21)(8,21)(10,1)(12,26)(14,2)(16,35)(18,15)
            (20,8)(22,5)(24,56)(26,1)(28,5)(30,15)(32,57)(34,1)(36,24)(38,2)
            (40,27)(42,11)(44,4)(46,3)(48,70)(50,8)(52,12)(54,8)(56,21)(58,5)
            (60,18)(62,8)(64,70)(66,9)(68,11)(70,10)(72,30)(74,1)(76,8)(78,2)
            (80,32)(82,3)(84,6)(86,1)(88,5)(90,17)(92,7)(94,3)(96,40)(100,4)
            (104,3)(106,1)(108,8)(112,8)(116,1)(120,16)(122,1)(124,2)(128,12)
            (132,9)(140,1)(144,1)(150,2)(160,3)(180,2)(192,12)(240,1)(384,1)
          };
        \end{axis}
      \end{tikzpicture}
      \caption{Number of Non-Abelian Singlets - $L1O3$ Extension}
      \label{figure:nahe_variations_na_singlets_l1o3}
    \end{subfigure}
    \caption{Statistics for the full NAHE variation extension data set.}
    \label{figure:nahe_variations}
\end{figure}

While it may seem from \autoref{table:nahe_variations_l1o3_groups} that the
order-$3$ models are more prone to enhancements,
\autoref{figure:nahe_variations_group_factors} makes it clear that is not the
case. The distribution of the number of gauge group factors for a model peaks
between $9$ and $11$ factors, as opposed to the peak at $8$ factors for the
order-$2$ models. However, there are several models with enhancements, even some
models with as few as $2$ distinct gauge group factors in them, something not
seen with the order-$2$ models. This implies there is a class of order-$3$ basis
vectors that greatly enhances the gauge group symmetries, while most order-$3$
models break them.

The number of $U(1)$ gauge groups per model is plotted in
\autoref{figure:nahe_variations_u1_factors}. The distribution of $U(1)$ peaks
between $5$ and $7$. More interestingly, a nontrivial number of models do not
have $U(1)$ symmetries at all. This implies, when combined with
\autoref{figure:nahe_variations_group_factors}, that in some models the $U(1)s$
are enhancing to larger (but still small relative to $SO(22)$ and $E_6$) gauge
groups. The mechanism producing this effect warrants further study, as it could
be used to reduce the number of $U(1)$ factors for order--layer combinations
that tend to produce too many $U(1)$s. The frequency of the GUT groups is
presented in \autoref{table:nahe_variations_gut_groups}.

The number of ST SUSYs is presented in
\autoref{figure:nahe_variations_st_susys}. While there are a statistically
significant number of enhanced ST SUSYs (expected from models with odd-ordered
right movers), the majority of these models have $N=0$ ST SUSY.

The number of non-Abelian singlets is plotted in
\autoref{figure:nahe_variations_na_singlets_l1o3}. The distribution of
non-Abelian singlets indicates that a large number of models do not have any
non-Abelian singlets. It is possible that this is related to the number of
models with no $U(1)$ factors.

\section{Models with GUT Groups}\label{section:gut_models}

As a parallel to the NAHE extension study, the subsets of models containing the
GUT groups $E_6$, $SO(10)$, $SU(5)\otimes U(1)$, $SU(4)\otimes SU(2)\otimes
SU(2)$ (Pati-Salam), $SU(3)\otimes SU(2)\otimes SU(2)$ (Left-Right Symmetric),
and $SU(3)\otimes SU(2)\otimes U(1)$ (MSSM) are examined. Like the NAHE study,
the usual statistics will be reported along with the number of net chiral
generations for models containing the GUT groups in question. If there is more
than one way to configure an observable sector, each configuration will be
counted when tallying the charged exotics and net chiral generations. For
example, a model may have two $E_6$ groups with different matter
representations. Each one would be counted individually when examining the
number of charged exotics and net chiral generations.

In order to calculated the net number of chiral fermion generations we utilize
the following expressions:

\begin{equation*}
\begin{array}{ll}
  E_6                                       & \abs{N_{27}-N_{\overline{27}}} \\
  SO(10)                                    & \abs{N_{16}-N_{\overline{16}}} \\
  SU(5) \otimes U(1)                        & \abs{\min (N_{10},N_{\overline{5}}) - \min (N_{\overline{10}},N_{5})} \\
  \textrm{Pati-Salam}                       & \abs{N_{(4,2,1)}-N_{(\overline{4},2,1)}} \\
  \textrm{Left-Right Symmetric\hspace{1cm}} & \abs{N_{(3,2,1)}-N_{(\overline{3},2,1)}} \\
  \textrm{MSSM}                             & \abs{N_{(3,2)}-N_{(\overline{3},2)}},~ \abs{N_{(3,1)}-N_{(\overline{3},1)}}.
\end{array}
\end{equation*}

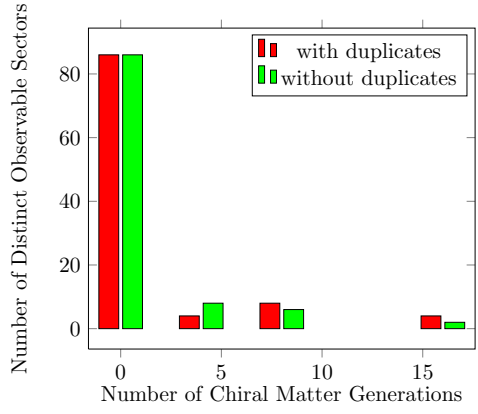
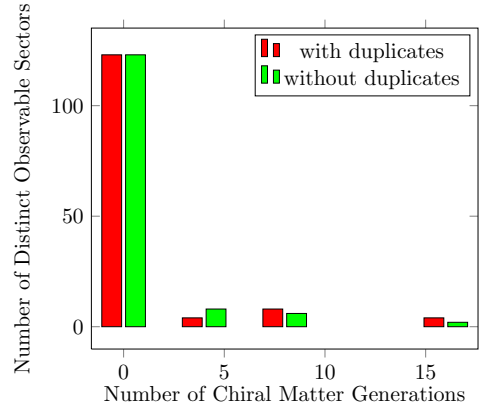
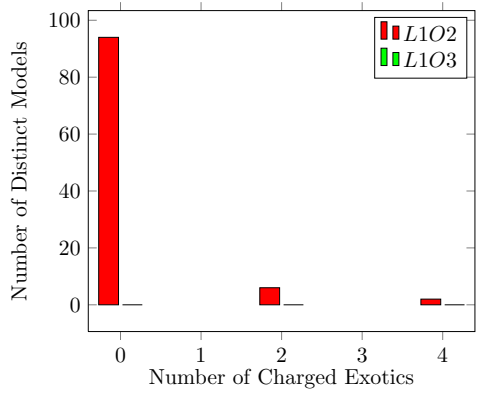
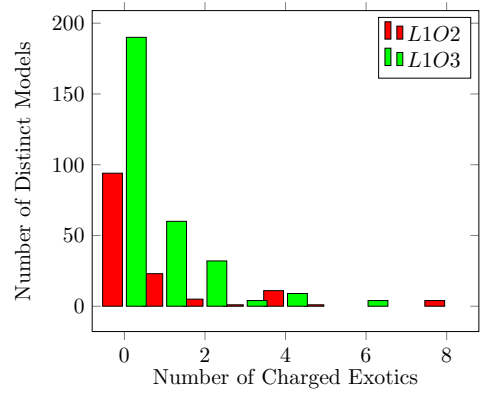
\begin{figure}[!ht]
  \centering
    \begin{subfigure}[b]{0.4\textwidth}
      \begin{tikzpicture}[scale=0.75]
        \begin{axis} [ybar, ylabel = Number of Distinct Observable Sectors,
                      xlabel = Number of Chiral Matter Generations]
          \addplot[draw=black, fill=red]coordinates{
            (0,86)(4,4)(8,8)(16,4)
          };
          \addlegendentry{with duplicates}
          \addplot[draw=black, fill=green]coordinates{
            (0,86)(4,8)(8,6)(16,2)
          };
          \addlegendentry{without duplicates}
        \end{axis}
      \end{tikzpicture}
      \caption{$E_6$ - Chiral Matter\\Generations}
      \label{figure:nahe_variations_e6_chiral_matter}
    \end{subfigure}
    \hspace{0.1\textwidth}
    \begin{subfigure}[b]{0.4\textwidth}
      \begin{tikzpicture}[scale=0.75]
        \begin{axis} [ybar, ylabel = Number of Distinct Observable Sectors,
                      xlabel = Number of Chiral Matter Generations]
          \addplot[draw=black, fill=red]coordinates{
            (0,123)(4,4)(8,8)(16,4)
          };
          \addlegendentry{with duplicates}
          \addplot[draw=black, fill=green]coordinates{
            (0,123)(4,8)(8,6)(16,2)
           };
          \addlegendentry{without duplicates}
        \end{axis}
      \end{tikzpicture}
      \caption{$SO(10)$ - Chiral Matter Generations}
      \label{figure:nahe_variations_so10_chiral_matter}
    \end{subfigure}
    \vspace{5mm}\\
    \begin{subfigure}[b]{0.4\textwidth}
      \begin{tikzpicture}[scale=0.75]
        \begin{axis} [ybar, ylabel = Number of Distinct Models,
                      xlabel = Number of Charged Exotics]
          \addplot[draw=black, fill=red]coordinates{
            (0,94)(2,6)(4,2)
          };
          \addlegendentry{$L1O2$}
          \addplot[draw=black, fill=green]coordinates{
            (0,0)(2,0)(4,0)
          };
          \addlegendentry{$L1O3$}
        \end{axis}
      \end{tikzpicture}
      \caption{$E_6$ - Charged Exotics}
      \label{figure:nahe_variations_e6_charged_exotics}
    \end{subfigure}
    \hspace{0.1\textwidth}
    \begin{subfigure}[b]{0.4\textwidth}
      \begin{tikzpicture}[scale=0.75]
        \begin{axis} [ybar, ylabel = Number of Distinct Models,
                      xlabel = Number of Charged Exotics]
          \addplot[draw=black, fill=red]coordinates{
            (0,94)(1,23)(2,5)(3,1)(4,11)(5,1)(8,4)
          };
          \addlegendentry{$L1O2$}
          \addplot[draw=black, fill=green]coordinates{
            (0,190)(1,60)(2,32)(3,4)(4,9)(6,4)
          };
          \addlegendentry{$L1O3$}
        \end{axis}
      \end{tikzpicture}
      \caption{$SO(10)$ - Charged Exotics}
      \label{figure:nahe_variations_so10_charged_exotics}
    \end{subfigure}
    \caption{The number of chiral matter generations and charged exotics for
    $E_6$ and $SO(10)$ models in the NAHE variation extensions.}
    \label{figure:nahe_variations_e6_so10_generations}
\end{figure}

\begin{figure}
  \centering
    \begin{subfigure}[b]{0.4\textwidth}
      \begin{tikzpicture}[scale=0.75]
        \begin{axis} [ybar, ylabel = Number of Distinct Models,
                      xlabel = Number of Charged Exotics, bar width = 3pt]
          \addplot[draw=red, fill=red]coordinates{
            (0,50)(2,38)(4,19)(8,26)(12,15)(16,19)(18,3)(20,2)(24,3)(32,2)
          };
        \end{axis}
      \end{tikzpicture}
      \caption{$SU(5) \otimes U(1)$}
    \end{subfigure}
    \hspace{0.1\textwidth}
    \begin{subfigure}[b]{0.4\textwidth}
      \begin{tikzpicture}[scale=0.75]
        \begin{axis} [ybar, ylabel = Number of Distinct Models,
                      xlabel = Number of Charged Exotics, bar width = 3pt]
            \addplot[draw=red, fill=red]coordinates{
              (0,42)(1,2)(2,20)(3,12)(4,10)(5,8)(6,2)(8,24)(9,6)(10,2)(12,36)
              (13,2)(14,22)(15,8)(16,60)(17,2)(18,8)(19,16)(20,46)(21,28)(22,30)
              (23,28)(24,58)(25,4)(26,14)(28,14)(29,4)(30,10)(31,16)(32,18)
              (33,10)(34,18)(35,14)(36,24)(37,2)(38,10)(39,16)(40,30)(41,2)
              (42,10)(43,6)(44,2)(45,8)(46,16)(47,12)(48,32)(49,6)(50,6)(51,4)
              (53,4)(54,2)(58,2)(62,16)(63,2)(64,38)(66,8)(69,8)(70,20)(71,12)
              (72,32)(73,8)(74,4)(77,8)(80,8)(88,8)(90,2)(95,2)(96,2)(138,4)
            };
          \end{axis}
      \end{tikzpicture}
      \caption{Pati-Salam}
    \end{subfigure}
    \vspace{5mm}\\
    \begin{subfigure}[b]{0.4\textwidth}
      \begin{tikzpicture}[scale=0.75]
        \begin{axis} [ybar, ylabel = Number of Distinct Models,
                      xlabel = Number of Charged Exotics, bar width = 3pt]
          \addplot[draw=red, fill=red]coordinates{
            (0,34)(2,18)(4,6)(5,4)(6,2)(7,4)(8,8)(9,2)(11,4)(12,6)(15,4)(16,2)
            (17,2)(19,2)(20,8)(21,2)(24,26)(25,2)(27,4)(28,20)(29,6)(30,4)(31,4)
            (32,28)(33,18)(34,8)(36,2)(37,4)(40,12)(41,4)(44,8)(45,6)(47,12)
            (48,20)(49,8)(65,4)
          };
        \end{axis}
      \end{tikzpicture}
      \caption{Left-Right Symmetric}
    \end{subfigure}
    \hspace{0.1\textwidth}
    \begin{subfigure}[b]{0.4\textwidth}
      \begin{tikzpicture}[scale=0.75]
        \begin{axis} [ybar, ylabel = Number of Distinct Models,
                      xlabel = Number of Charged Exotics, bar width = 3pt]
          \addplot[draw=red, fill=red]coordinates{
            (4,2)(8,2)(16,12)(18,1)(20,10)(21,4)(22,11)(24,24)(27,6)(28,2)
            (29,1)(31,2)(32,19)(33,10)(34,5)(35,6)(36,12)(37,1)(38,1)(40,15)
            (41,2)(42,1)(44,4)(48,2)(49,2)(60,4)(64,4)(65,1)(67,1)
          };
        \end{axis}
      \end{tikzpicture}
      \caption{MSSM}
    \end{subfigure}
    \caption{The number of charged exotics for $SU(5) \otimes U(1)$, Pati-Salam,
    Left-Right Symmetric and MSSM-like models in the NAHE variation extensions.}
    \label{figure:nahe_variations_fsu5_ps_lrs_mssm_matter}
\end{figure}
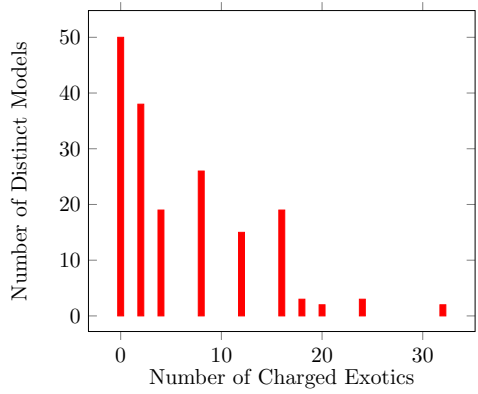
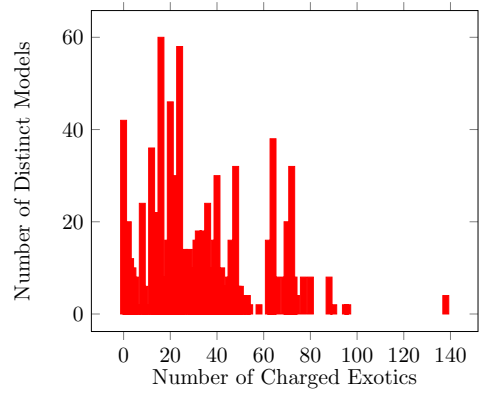
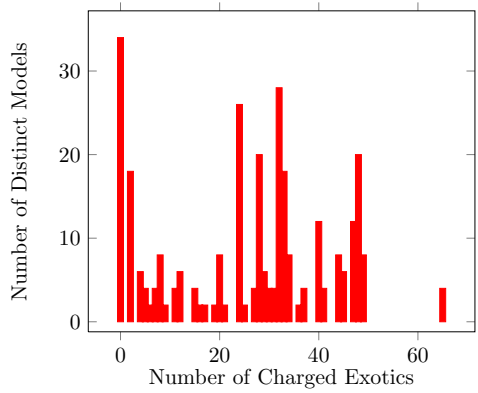
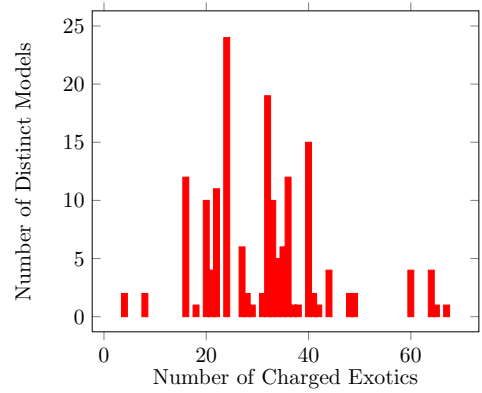

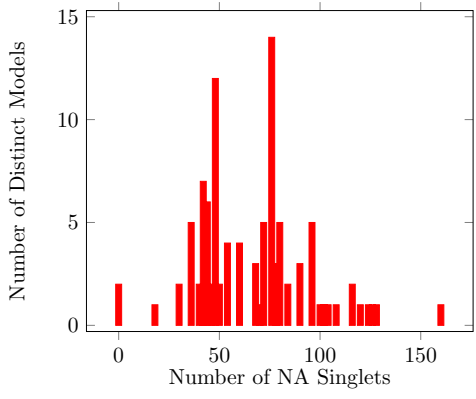
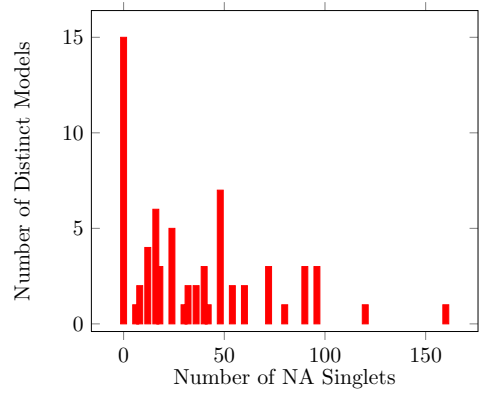
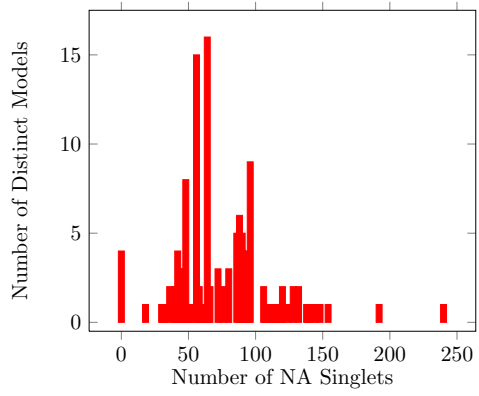
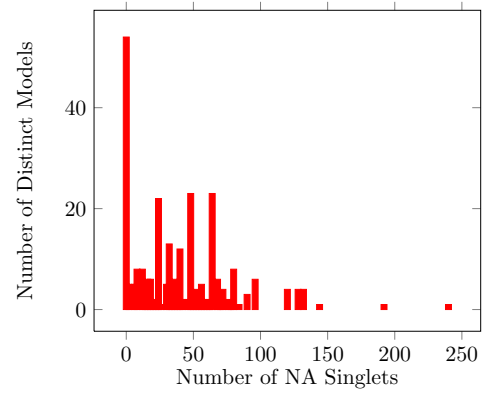
\begin{figure}[!ht]
  \centering
    \begin{subfigure}[b]{0.4\textwidth}
      \begin{tikzpicture}[scale=0.75]
        \begin{axis} [ybar, ylabel = Number of Distinct Models,
                      xlabel = Number of NA Singlets, bar width = 3pt]
          \addplot[draw=red, fill=red]coordinates{
            (0,2)(18,1)(30,2)(36,5)(40,2)(42,7)(44,6)(46,2)(48,12)(50,2)(54,4)
            (60,4)(68,3)(70,1)(72,5)(76,14)(78,3)(80,5)(84,2)(90,3)(96,5)(100,1)
            (102,1)(104,1)(108,1)(116,2)(120,1)(124,1)(126,1)(128,1)(160,1)
          };
        \end{axis}
      \end{tikzpicture}
      \caption{$E_6$ - Number of NA Singlets ($L1O2$)}
      \label{figure:nahe_variations_e6_l1o2_na_singlets}
    \end{subfigure}
    \hspace{0.1\textwidth}
    \begin{subfigure}[b]{0.4\textwidth}
      \begin{tikzpicture}[scale=0.75]
        \begin{axis} [ybar, ylabel = Number of Distinct Models,
                      xlabel = Number of NA Singlets, bar width = 3pt]
          \addplot[draw=red, fill=red]coordinates{
            (0,15)(6,1)(8,2)(12,4)(16,6)(18,3)(24,5)(30,1)(32,2)(36,2)(40,3)
            (42,1)(48,7)(54,2)(60,2)(72,3)(80,1)(90,3)(96,3)(120,1)(160,1)
          };
        \end{axis}
      \end{tikzpicture}
      \caption{$E_6$ - Number of NA Singlets ($L1O3$)}
      \label{figure:nahe_variations_e6_l1o3_na_singlets}
    \end{subfigure}
    \vspace{5mm}\\
    \begin{subfigure}[b]{0.4\textwidth}
      \begin{tikzpicture}[scale=0.75]
        \begin{axis} [ybar, ylabel = Number of Distinct Models,
                      xlabel = Number of NA Singlets, bar width = 3pt]
          \addplot[draw=red, fill=red]coordinates{
            (0,4)(18,1)(30,1)(32,1)(36,2)(40,2)(42,4)(44,3)(46,1)(48,8)(50,1)
            (54,1)(56,15)(58,2)(60,1)(64,16)(66,2)(72,3)(76,2)(80,3)(86,5)
            (88,6)(90,5)(92,4)(94,2)(96,9)(106,2)(108,1)(112,1)(114,1)(116,1)
            (118,1)(120,2)(124,1)(128,2)(132,2)(138,1)(140,1)(144,1)(148,1)
            (154,1)(192,1)(240,1)
          };
        \end{axis}
      \end{tikzpicture}
      \caption{$SO(10)$ - Number of NA Singlets ($L1O2$)}
      \label{figure:nahe_variations_so10_l1o2_na_singlets}
    \end{subfigure}
    \hspace{0.1\textwidth}
    \begin{subfigure}[b]{0.4\textwidth}
      \begin{tikzpicture}[scale=0.75]
        \begin{axis} [ybar, ylabel = Number of Distinct Models,
                      xlabel = Number of NA Singlets, bar width = 3pt]
          \addplot[draw=red, fill=red]coordinates{
            (0,54)(2,5)(6,5)(8,8)(12,8)(14,1)(16,6)(18,6)(20,2)(22,1)(24,22)
            (26,1)(30,5)(32,13)(34,1)(36,6)(38,1)(40,12)(42,2)(46,2)(48,23)
            (50,2)(52,4)(54,1)(56,5)(58,2)(60,2)(62,1)(64,23)(68,6)(70,1)(72,4)
            (76,2)(78,1)(80,8)(84,1)(90,3)(96,6)(120,4)(128,4)(132,4)(144,1)
            (192,1)(240,1)
          };
        \end{axis}
      \end{tikzpicture}
      \caption{$SO(10)$ - Number of NA Singlets ($L1O3$)}
      \label{figure:nahe_variations_so10_l1o3_na_singlets}
    \end{subfigure}
    \caption{NA Singlet statistics for the $E_6$ and $SO(10)$ models in the NAHE variation extenstions data set.}
    \label{figure:nahe_variations_e6_so10_na_singlets}
\end{figure}

\begin{figure}
  \centering
    \begin{subfigure}[b]{0.4\textwidth}
      \begin{tikzpicture}[scale=0.75]
        \begin{axis} [ybar, ylabel = Number of Distinct Models,
                      xlabel = NA Singlets, bar width = 3pt]
          \addplot[draw=red, fill=red]coordinates{
            (0,7)(6,1)(8,1)(10,1)(12,1)(14,1)(16,4)(18,1)(20,1)(22,2)(24,12)
            (28,1)(30,4)(32,9)(36,5)(40,4)(42,1)(44,2)(46,1)(48,11)(50,2)(52,3)
            (56,5)(58,3)(60,5)(62,2)(64,12)(66,4)(68,3)(70,4)(72,4)(74,1)(76,3)
            (78,1)(80,8)(82,3)(84,4)(86,1)(88,3)(90,3)(92,4)(94,3)(96,3)(106,1)
            (108,3)(112,3)(120,1)(122,1)(124,1)(128,1)
          };
        \end{axis}
      \end{tikzpicture}
      \caption{$SU(5) \otimes U(1)$}
      \label{}
    \end{subfigure}
    \hspace{0.1\textwidth}
    \begin{subfigure}[b]{0.4\textwidth}
      \begin{tikzpicture}[scale=0.75]
        \begin{axis} [ybar, ylabel = Number of Distinct Models,
                      xlabel = NA Singlets, bar width = 3pt]
          \addplot[draw=red, fill=red]coordinates{
            (0,33)(2,3)(6,3)(8,3)(12,6)(14,1)(16,4)(18,2)(20,3)(24,15)(28,2)
            (32,12)(36,5)(40,5)(42,2)(48,4)(52,1)(60,1)(64,7)(68,1)(80,1)(84,1)
            (96,3)(104,1)(108,2)(112,1)(120,1)(192,2)
          };
        \end{axis}
      \end{tikzpicture}
      \caption{Pati-Salam}
    \end{subfigure}
    \vspace{5mm}\\
    \begin{subfigure}[b]{0.4\textwidth}
      \begin{tikzpicture}[scale=0.75]
        \begin{axis} [ybar, ylabel = Number of Distinct Models,
                      xlabel = NA Singlets, bar width = 3pt]
          \addplot[draw=red, fill=red]coordinates{
            (0,3)(6,2)(22,1)(24,3)(32,8)(36,2)(38,1)(40,3)(42,5)(46,1)(48,11)
            (52,2)(54,1)(62,3)(64,4)(68,1)(72,5)(80,1)(90,2)(96,2)
          };
        \end{axis}
      \end{tikzpicture}
      \caption{Left-Right Symmetric}
    \end{subfigure}
    \hspace{0.1\textwidth}
    \begin{subfigure}[b]{0.4\textwidth}
      \begin{tikzpicture}[scale=0.75]
        \begin{axis} [ybar, ylabel = Number of Distinct Models,
                      xlabel = NA Singlets, bar width = 3pt]
          \addplot[draw=red, fill=red] coordinates{
            (0,3)(6,3)(8,1)(22,1)(24,3)(32,8)(36,2)(38,1)(40,3)(42,5)(46,1)
            (48,11)(52,2)(54,1)(62,3)(64,4)(68,1)(72,5)(80,1)(90,2)(96,2)
          };
        \end{axis}
      \end{tikzpicture}
      \caption{MSSM}
    \end{subfigure}
    \caption{NA Singlet statistics for the $SU(5) \otimes U(1)$, Pati-Salam,
    Left-Right Symmetric, and MSSM-like models. Note that these only arise as
    $L1O3$ extensions.}
    \label{figure:nahe_variations_fsu5_ps_lrs_mssm_na_singlets}
\end{figure}
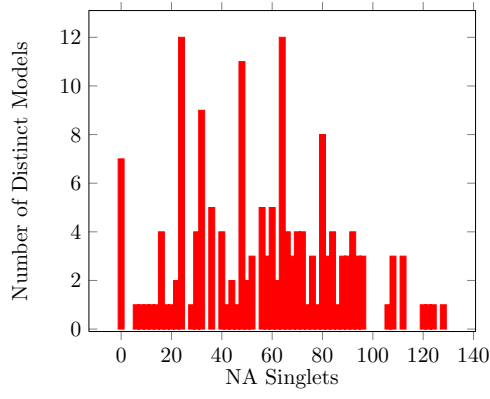
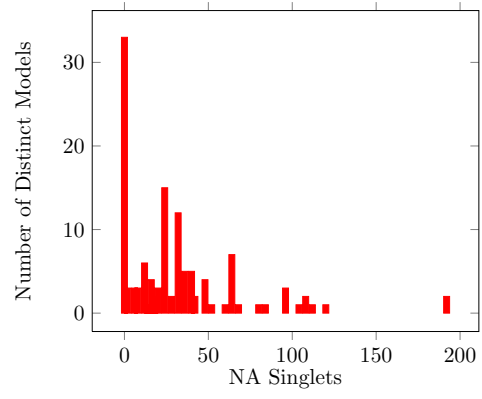
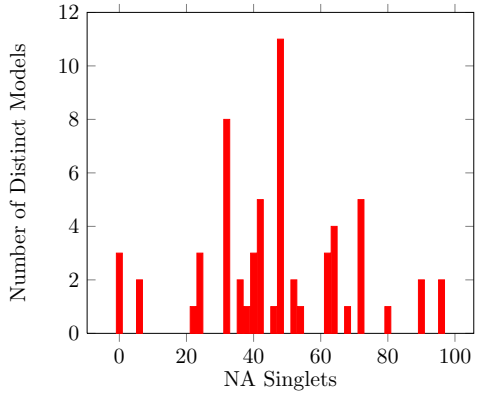
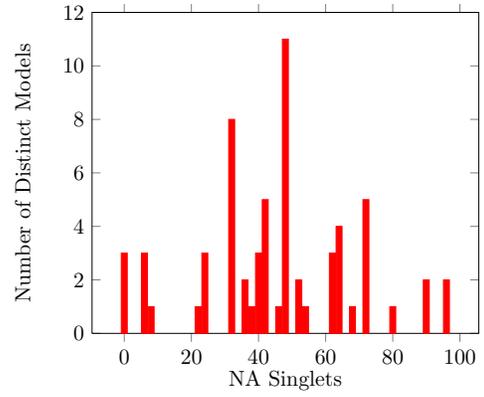

Upon analysis it's found that the $L1O2$ extensions yield $E_6$ and $SO(10)$
observable sectors with net chiral generations while no models with $SU(5)
\otimes U(1)$, $SO(6) \otimes SO(4)$, $SU(3) \otimes SU(2) \otimes SU(2)$, nor
$SU(3) \otimes SU(2) \otimes U(1)$ have this property. This is a consequence of
the fact that the latter groups only arise from $L1O3$ extensions which are not
conducive to production of net chiral generations. The distribution of net
chiral generations, as well as charged exotic matter, by gauge group is provided
in \autoref{figure:nahe_variations_e6_so10_generations} and
\autoref{figure:nahe_variations_fsu5_ps_lrs_mssm_matter}. The distributions of
number of non-Abelian singlets, by gauge group, can be found in
\autoref{figure:nahe_variations_e6_so10_na_singlets} and
\autoref{figure:nahe_variations_fsu5_ps_lrs_mssm_na_singlets}.

In addition to matter content, the hidden sector gauge content is tabulated for
each of the aforementioned gauge groups:
\autoref{table:nahe_variations_e6_groups},
\autoref{table:nahe_variations_so10_groups} and
\autoref{table:nahe_variations_gut_hidden_sector}. We can see from
\autoref{table:nahe_variations_gut_groups} that the NAHE variation extensions
favor $E_6$ and $SO(10)$ over the other groups. This is easily understood as
$E_6$ is already present and the breaking $E_6$ to $SO(10)$ is rather straight
forward. However, in order to produce the low-rank $SU(n + 1)$ groups either the
$U(1)$s must be enhanced or there must be significant breaking of either the
$E_6$ or $SO(22)$, neither of which readily occur with a single layer or at low
order.

\subsection{ST SUSYs}\label{section:susy}
\begin{figure}[!ht]
  \centering
    \begin{subfigure}[b]{0.4\textwidth}
      \begin{tikzpicture}[scale=0.75]
        \begin{axis} [ybar stacked, ylabel = Number of Distinct Models,
                      xlabel = Number of ST SUSYs]
          \addplot[draw=black, fill=red]coordinates{
            (0,52) (1,49) (2, 0) (4, 0)
          };
          \addlegendentry{$L1O2$}
          \addplot[draw=black, fill=green]coordinates{
            (0,44) (1,15) (2,9) (4, 0)
          };
          \addlegendentry{$L1O3$}
        \end{axis}
      \end{tikzpicture}
      \caption{$E_6$ - Number of ST SUSYs}
      \label{figure:nahe_variations_e6_st_susys}
    \end{subfigure}
    \hspace{0.1\textwidth}
    \begin{subfigure}[b]{0.4\textwidth}
      \begin{tikzpicture}[scale=0.75]
        \begin{axis} [ybar stacked, ylabel = Number of Distinct Models,
                      xlabel = Number of ST SUSYs]
          \addplot[draw=black, fill=red]coordinates{
            (0,66) (1,59) (2, 0) (4, 0)
          };
          \addlegendentry{$L1O2$}
          \addplot[draw=black, fill=green]coordinates{
            (0,216) (1,44) (2,11) (4, 0)
          };
          \addlegendentry{$L1O3$}
        \end{axis}
      \end{tikzpicture}
      \caption{$SO(10)$ - Number of ST SUSYs}
      \label{figure:nahe_variations_so10_st_susys}
    \end{subfigure}
    \vspace{5mm}\\
    \begin{subfigure}[b]{0.4\textwidth}
      \begin{tikzpicture}[scale=0.75]
        \begin{axis} [ybar stacked, ylabel = Number of Distinct Models,
                      xlabel = Number of ST SUSYs,, legend pos = north east]
          \addplot[draw=black, fill=red]coordinates{
            (0,146) (1,19) (2, 0) (4, 0)
          };
          \addlegendentry{$SU(5)\otimes U(1)$}
          \addplot[draw=black, fill=green]coordinates{
            (0,98) (1,22) (2,5) (4, 0)
          };
          \addlegendentry{Pati-Salam}
          \addplot[draw=black, fill=blue]coordinates{
            (0,57) (1,4) (2, 0) (4, 0)
          };
          \addlegendentry{$LRS$}
          \addplot[draw=black, fill=yellow]coordinates{
            (0,57) (1,5) (2,1) (4, 0)
          };
          \addlegendentry{$MSSM$}
        \end{axis}
      \end{tikzpicture}
      \caption{Number of ST SUSYs}
      \label{figure:nahe_variations_fsu5_ps_lrs_mssm_st_susys}
    \end{subfigure}
    \caption{ST SUSY statistics for Various GUT models. Note that only the $E_6$
    and $SO(10)$ occur from have $L1O2$ extensions.}
    \label{figure:nahe_variations_st_susys_by_group}
\end{figure}
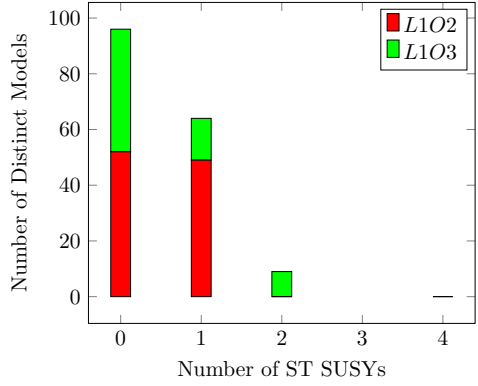
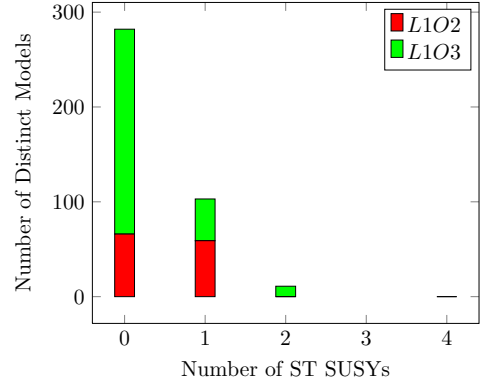
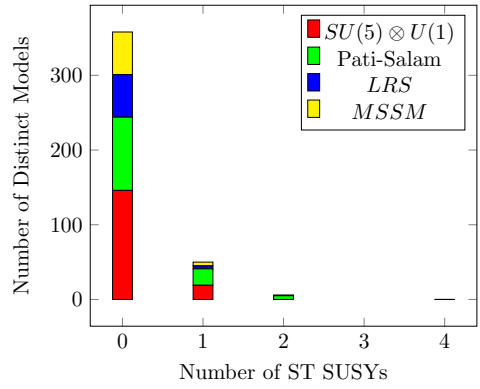

The distributions of ST SUSYs for the entire data set can be found in
\autoref{figure:nahe_variations_st_susys} with a breakdown by gauge group in
\autoref{figure:nahe_variations_st_susys_by_group}.

\begin{figure}
  \centering
    \includegraphics[scale=0.85]{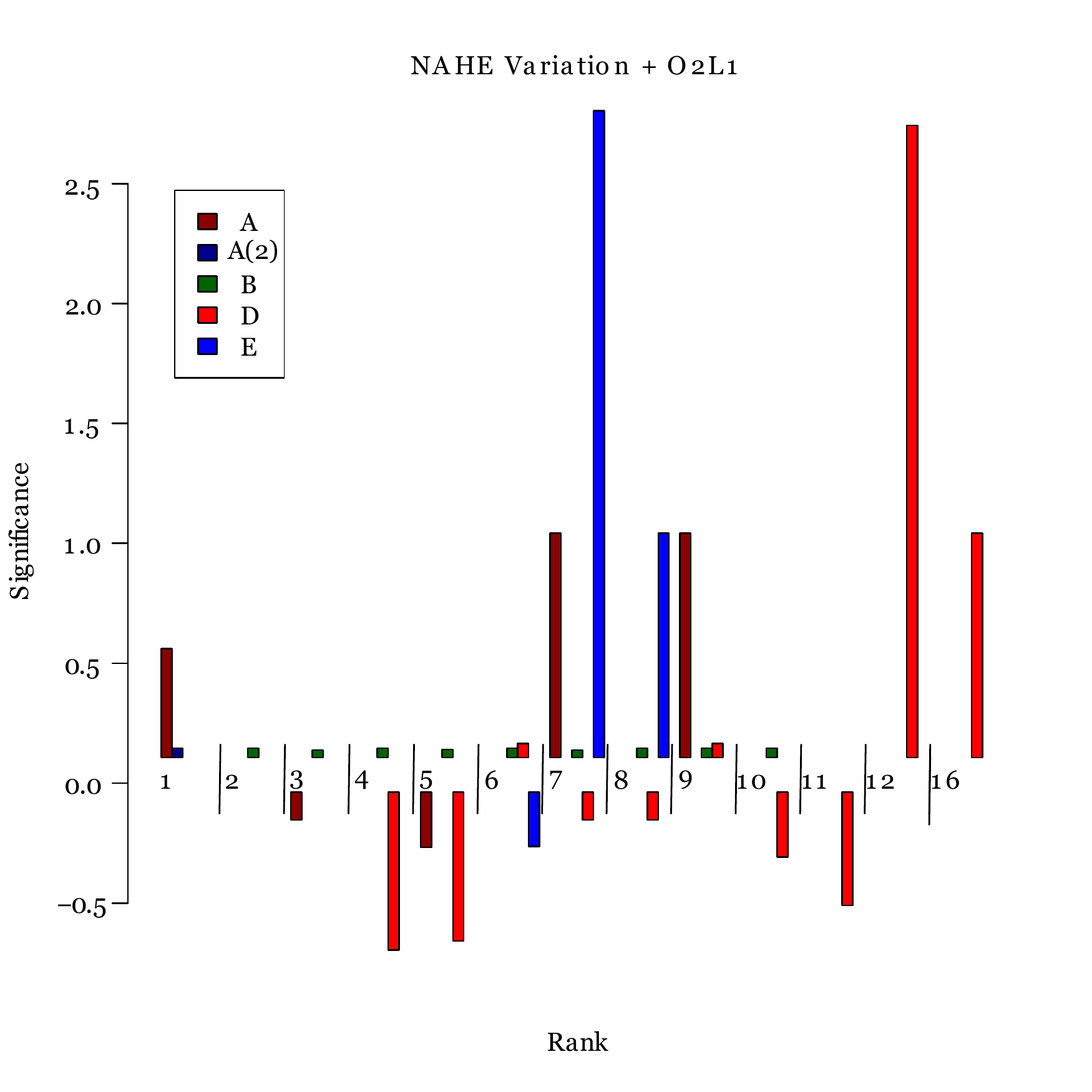}
    \caption{The significance values for models in the NAHE variation $L1O2$ extensions with regard to ST SUSY. Any (absolute) significance values
    greater than three indicate a strong statistical significance.}
    \label{figure:nahe_variations_l1o2_st_susy_significances}
\end{figure}

\begin{figure}
  \centering
    \includegraphics[scale=0.85]{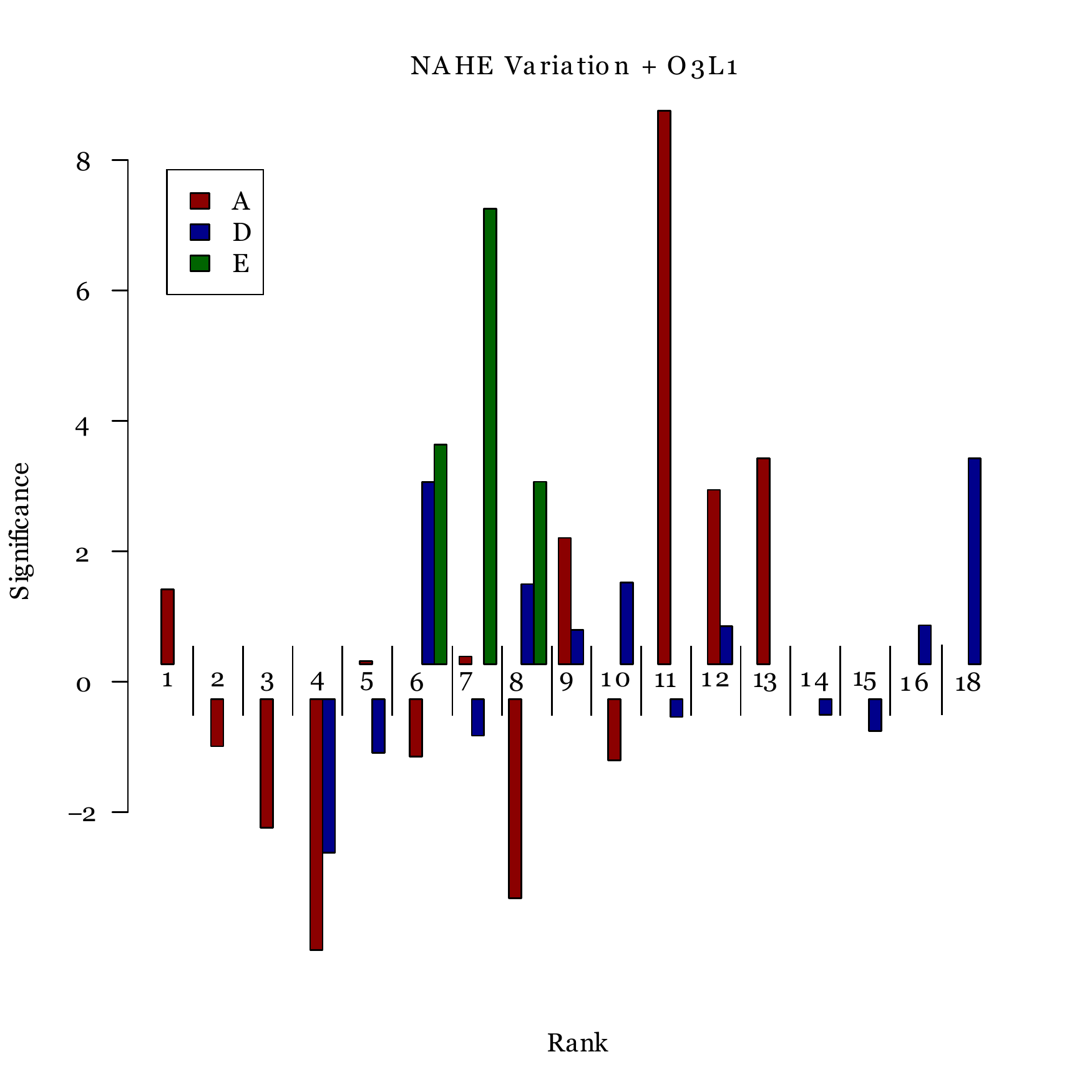}
    \caption{The significance values for models in the NAHE variation $L1O3$
    extensions. Any (absolute) significance values greater than three indicate a
    strong statistical significance.}
    \label{figure:nahe_variations_l1o3_st_susy_significances}
\end{figure}

The $L1O2$ models all have the same distributions regardless of which GUT is
chosen. In these models, the gauge content does not statistically couple to the
ST SUSY. For the $L1O3$ models, however, some of the GUT groups do appear to
have such a coupling. In particular, the occurence of $E_6$ models $N=2$ ST SUSY
is disproportionately high while $SU(5)\otimes U(1)$, Left-Right Symmetric, and
MSSM models with $N=1$ ST SUSY have a reduced occurence. As all of the models
containing these GUTs have at least a single $U(1)$, there could be a
correlation between the number of $U(1)$s and the number of ST SUSYs. Further
investigations of these findings show several statistical couplings for higher
ST SUSY models containing certain gauge group factors. The methodology used to
analyze these couplings was detailed in \cite{Renner:2011}. The observed
significances are plotted in
\autoref{figure:nahe_variations_l1o2_st_susy_significances} and
\autoref{figure:nahe_variations_l1o3_st_susy_significances} for the $L1O2$ and
$L1O3$ NAHE variation extensions, respectively.

While there are no significant gauge groups in the $L1O2$ extensions, several
groups are significant with regard to enhanced ST SUSYs in the NAHE $L1O3$
extensions. In particular, the three exceptional groups, as well as $SO(12)$,
$SU(12)$, $SU(13)$, $SU(14)$, and $SO(36)$ all have a significant statistical
correlation with the average number of ST SUSYs. This is likely due to the
additional basis vector adding a gravitino generating sector, which is common
with odd-order extensions, and additional roots for the gauge groups. Further
analysis will be needed to confirm the cause of this significance. It is also
worth noting that one group, $SU(5)$, has a negative impact on ST SUSYs. If
this trend occurs for more odd-ordered extensions of the NAHE variation, it may
affect the viability of realistic flipped-$SU(5)$ models derived from this
variation.

\section{Models with Mirroring}\label{section:mirroring}
The larger sets of matching boundary conditions, seen in
\autoref{table:nahe_variation_basis}, are expected to lead to models with
mirrored gauge groups and matter states. Only one model, generated by
\autoref{table:mirrored_model_basis_vectors}, in those discussed thus far
exhibits full gauge mirroring. However, the matter states are not mirrored. The
particle content of that model is presented in
\autoref{table:mirrored_model_content}.

The gauge groups are completely mirrored, and the matter representations are
almost mirrored between one another. There is a state charged as a $\bm{16}$
under both $SO(16)$ groups and one charged as a $\bm{128}$ under one of the
$SO(16)$ groups, but not the other. Thus, the matter is not mirrored. The
potential for mirroring is clear from the basis vectors:
$\bar{\psi}^{1,\dots,5}$ and $\bar{\eta}^{1,2,3}$ are mirrored with
$\bar{\phi}^{1,\dots,8}$.There are also many models in which the observable and
some of the hidden matter is mirrored, but include a shadow sector gauge group
whose matter representations are not coupled.

These have been presented and discussed in \cite{Greenwald:2009}.

\section{Conclusions}\label{section:conclusion}
Though there were many models containing GUTs in the data sets explored in this
study, a vast majority of them do not contain any net chiral fermion
generations. No three-generation models were found. These conclusions are
summarized in \autoref{table:nahe_variations_gut_generations}.

While there were more models with GUT gauge groups in the NAHE variation $L1O3$
extensions, none of them had any net chiral matter generations, implying that the
added basis vector produces the barred and unbarred generations in even pairs,
if at all. More complicated basis vector sets will need to be studied to
determine if any NAHE variation based quasi-realistic models can be constructed.

The distribution of ST SUSYs across the subsets of GUT models was also
examined. It was concluded that, as was the case with the NAHE study, $E_6$ has
a statistical coupling to enhanced ST SUSYs for order-3 models. Additionally,
data sets in which all of the models contained at least one $U(1)$ factor with a
GUT group had fewer models with $N=1$ ST SUSY.

Models with partial gauge group mirroring were also discussed, with a model
presented that has complete gauge group mirroring. While a statistical search
algorithm for finding quasi-mirrored models has not yet been completed, it will
be used in future work to examine models with this property.
\begin{table}
  \centering
    \caption{A summary of the GUT group study with regard to the number of chiral fermion generations in the NAHE variation investigation.}
    \begin{tabular}{||c|c|c||}
    \hline \hline
    GUT                       & Net Chiral Generations?   & Three Generations?\\
    \hline \hline
    $L1O2 ~E_6$               & Yes                       & No                \\
    \hline
    $L1O2 ~SO(10)$            & Yes                       & No                \\
    \hline
    $L1O3 ~E_6$               & No                        & No                \\
    \hline
    $L1O3 ~SO(10)$            & No                        & No                \\
    \hline
    $L1O3 ~SU(5)\otimes U(1)$ & No                        & No                \\
    \hline
    $L1O3$ Pati-Salam         & No                        & No                \\
    \hline
    $L1O3$ L-R Symmetric      & No                        & No                \\
    \hline
    $L1O3$ MSSM               & No                        & No                \\
    \hline\hline
    \end{tabular}
    \label{table:nahe_variations_gut_generations}
\end{table}

\section{Acknowledgements}
This work was supported by funding from Baylor University.

\appendix
\section{Tables}
\begin{table}[!ht]
  \renewcommand{\arraystretch}{0.75}
  \caption{The GUT group content of the NAHE variation extenstions data set.}
  \label{table:nahe_variations_gut_groups}
  \begin{tabular}{||C{4.5cm}|C{2.75cm}|C{2.75cm}|C{2.75cm}|C{2.75cm}||}
    \hline\hline
    \multirow{2}{*}{GUT Group}
              & \multicolumn{2}{|c|}{$L1O2$}
              & \multicolumn{2}{|c||}{$L1O3$} \\ \cline{2-5}
              & Number of Unique Models & \% of Unique Models
              & Number of Unique Models & \% of Unique Models \\
    \hline\hline
    $E_6$                             & 101 & 32.69\% & 68  & 5.832\% \\
    \hline
    $SO(10)$                          & 125 & 40.45\% & 271 & 23.24\% \\
    \hline
    $SU(5)\otimes U(1)$               & 0   & 0\%     & 165 & 14.15\% \\
    \hline
    $SU(4)\otimes SU(2)\otimes SU(2)$ & 0   & 0\%     & 125 & 10.72\% \\
    \hline
    $SU(3)\otimes SU(2)\otimes SU(2)$ & 0   & 0\%     & 61  & 5.232\% \\
    \hline
    $SU(3)\otimes SU(2)\otimes U(1)$  & 0   & 0\%     & 63  & 5.403\% \\
    \hline \hline
  \end{tabular}
\end{table}

\begin{table}[!ht]
  \renewcommand{\arraystretch}{0.75}
  \caption{The gauge group content of the NAHE variation data set}
  \label{table:nahe_variations_groups}
  \begin{subtable}[t]{7cm}
    \caption{Layer 1, Order 2}
    \begin{tabular}{||C{2.5cm}|C{2cm}|C{2cm}||}
      \hline \hline
      Gauge Group & Number of Unique Models & \% of Unique Models \\
      \hline \hline
      $SU(2)$&131&42.39\%\\
      \hline
      $SU(2)^{(2)}$&18&5.825\%\\
      \hline
      $SU(4)$&33&10.68\%\\
      \hline
      $SU(6)$&99&32.04\%\\
      \hline
      $SU(8)$&1&0.3236\%\\
      \hline
      $SU(10)$&1&0.3236\%\\
      \hline
      $SO(5)$&18&5.825\%\\
      \hline
      $SO(7)$&12&3.883\%\\
      \hline
      $SO(9)$&18&5.825\%\\
      \hline
      $SO(11)$&14&4.531\%\\
      \hline
      $SO(13)$&18&5.825\%\\
      \hline
      $SO(15)$&12&3.883\%\\
      \hline
      $SO(17)$&18&5.825\%\\
      \hline
      $SO(19)$&18&5.825\%\\
      \hline
      $SO(21)$&18&5.825\%\\
      \hline
      $SO(8)$&30&9.709\%\\
      \hline
      $SO(10)$&125&40.45\%\\
      \hline
      $SO(12)$&38&12.3\%\\
      \hline
      $SO(14)$&33&10.68\%\\
      \hline
      $SO(16)$&33&10.68\%\\
      \hline
      $SO(18)$&38&12.3\%\\
      \hline
      $SO(20)$&36&11.65\%\\
      \hline
      $SO(22)$&31&10.03\%\\
      \hline
      $SO(24)$&2&0.6472\%\\
      \hline
      $SO(32)$&1&0.3236\%\\
      \hline
      $E_6$&101&32.69\%\\
      \hline
      $E_7$&3&0.9709\%\\
      \hline
      $E_8$&1&0.3236\%\\
      \hline
      $U(1)$&304&98.38\%\\
      \hline \hline
    \end{tabular}
    \label{table:nahe_variations_l1o2_groups}
  \end{subtable}
  \qquad
  \begin{subtable}[t]{7cm}
    \caption{Layer 1, Order 3}
    \begin{tabular}{||C{2.5cm}|C{2cm}|C{2cm}||}
      \hline\hline
      Gauge Group & Number of Unique Models & \% of Unique Models \\
      \hline\hline
      $SU(2)$&731&62.69\%\\
      \hline
      $SU(3)$&128&10.98\%\\
      \hline
      $SU(4)$&355&30.45\%\\
      \hline
      $SU(5)$&165&14.15\%\\
      \hline
      $SU(6)$&167&14.32\%\\
      \hline
      $SU(7)$&75&6.432\%\\
      \hline
      $SU(8)$&143&12.26\%\\
      \hline
      $SU(9)$&164&14.07\%\\
      \hline
      $SU(10)$&169&14.49\%\\
      \hline
      $SU(11)$&137&11.75\%\\
      \hline
      $SU(12)$&56&4.803\%\\
      \hline
      $SU(13)$&4&0.3431\%\\
      \hline
      $SU(14)$&1&0.08576\%\\
      \hline
      $SO(8)$&376&32.25\%\\
      \hline
      $SO(10)$&271&23.24\%\\
      \hline
      $SO(12)$&151&12.95\%\\
      \hline
      $SO(14)$&81&6.947\%\\
      \hline
      $SO(16)$&106&9.091\%\\
      \hline
      $SO(18)$&28&2.401\%\\
      \hline
      $SO(20)$&69&5.918\%\\
      \hline
      $SO(22)$&5&0.4288\%\\
      \hline
      $SO(24)$&11&0.9434\%\\
      \hline
      $SO(28)$&13&1.115\%\\
      \hline
      $SO(30)$&1&0.08576\%\\
      \hline
      $SO(32)$&2&0.1715\%\\
      \hline
      $SO(36)$&1&0.08576\%\\
      \hline
      $E_6$&68&5.832\%\\
      \hline
      $E_7$&24&2.058\%\\
      \hline
      $E_8$&9&0.7719\%\\
      \hline
      $U(1)$&1002&85.93\%\\
      \hline\hline
    \end{tabular}
    \label{table:nahe_variations_l1o3_groups}
  \end{subtable}
\end{table}

\begin{table}[!ht]
  \renewcommand{\arraystretch}{0.75}
  \caption{The hidden sector gauge group content for the NAHE variation
  extension models with $E_6$ observable.}
  \label{table:nahe_variations_e6_groups}
  \begin{subtable}[t]{7cm}
    \caption{Layer 1, Order 2}
    \label{table:nahe_variations_e6_l102_groups}
    \begin{tabular}{||C{2.5cm}|C{2cm}|C{2cm}||}
      \hline\hline
      Gauge Group & Number of Unique Models & \% of Unique Models \\
      \hline\hline
      $SU(2)$&14&13.86\%\\
      \hline
      $SU(2)^{(2)}$&8&7.921\%\\
      \hline
      $SU(4)$&10&9.901\%\\
      \hline
      $SO(5)$&6&5.941\%\\
      \hline
      $SO(7)$&2&1.98\%\\
      \hline
      $SO(9)$&6&5.941\%\\
      \hline
      $SO(11)$&6&5.941\%\\
      \hline
      $SO(13)$&6&5.941\%\\
      \hline
      $SO(15)$&2&1.98\%\\
      \hline
      $SO(17)$&6&5.941\%\\
      \hline
      $SO(19)$&8&7.921\%\\
      \hline
      $SO(21)$&6&5.941\%\\
      \hline
      $SO(8)$&8&7.921\%\\
      \hline
      $SO(10)$&14&13.86\%\\
      \hline
      $SO(12)$&14&13.86\%\\
      \hline
      $SO(14)$&9&8.911\%\\
      \hline
      $SO(16)$&9&8.911\%\\
      \hline
      $SO(18)$&14&13.86\%\\
      \hline
      $SO(20)$&12&11.88\%\\
      \hline
      $SO(22)$&8&7.921\%\\
      \hline
      $E_8$&1&0.9901\%\\
      \hline
      $U(1)$&101&100\%\\
      \hline \hline
    \end{tabular}
  \end{subtable}
  \qquad
  \begin{subtable}[t]{7cm}
    \caption{Layer 1, Order 3}
    \label{table:nahe_variations_e6_l103_groups}
    \begin{tabular}{||C{2.5cm}|C{2cm}|C{2cm}||}
      \hline \hline
      Gauge Group & Number of Unique Models & \% of Unique Models \\
      \hline \hline
      $SU(2)$&31&45.59\%\\
      \hline
      $SU(3)$&1&1.471\%\\
      \hline
      $SU(4)$&12&17.65\%\\
      \hline
      $SU(6)$&4&5.882\%\\
      \hline
      $SU(8)$&6&8.824\%\\
      \hline
      $SU(9)$&10&14.71\%\\
      \hline
      $SU(10)$&8&11.76\%\\
      \hline
      $SU(11)$&5&7.353\%\\
      \hline
      $SU(12)$&4&5.882\%\\
      \hline
      $SU(13)$&1&1.471\%\\
      \hline
      $SO(8)$&9&13.24\%\\
      \hline
      $SO(10)$&15&22.06\%\\
      \hline
      $SO(12)$&12&17.65\%\\
      \hline
      $SO(14)$&5&7.353\%\\
      \hline
      $SO(16)$&3&4.412\%\\
      \hline
      $SO(18)$&4&5.882\%\\
      \hline
      $SO(20)$&2&2.941\%\\
      \hline
      $SO(22)$&2&2.941\%\\
      \hline
      $E_8$&2&2.941\%\\
      \hline
      $U(1)$&68&100\%\\
      \hline \hline
    \end{tabular}
  \end{subtable}
\end{table}

\begin{table}[!ht]
  \renewcommand{\arraystretch}{0.75}
  \caption{The hidden sector gauge group content for the NAHE variation
  extension models with $SO(10)$ observable.}
  \label{table:nahe_variations_so10_groups}
  \begin{subtable}[t]{7cm}
    \caption{Layer 1, Order 2}
    \label{table:nahe_variations_so10_l1o2_groups}
    \begin{tabular}{||C{2.5cm}|C{2cm}|C{2cm}||}
      \hline\hline
      Gauge Group & Number of Unique Models & \% of Unique Models \\
      \hline \hline
      $SU(2)$&23&18.4\%\\
      \hline
      $SU(2)^{(2)}$&8&6.4\%\\
      \hline
      $SU(4)$&10&8\%\\
      \hline
      $SU(6)$&10&8\%\\
      \hline
      $SO(5)$&6&4.8\%\\
      \hline
      $SO(7)$&2&1.6\%\\
      \hline
      $SO(9)$&6&4.8\%\\
      \hline
      $SO(11)$&6&4.8\%\\
      \hline
      $SO(13)$&6&4.8\%\\
      \hline
      $SO(15)$&2&1.6\%\\
      \hline
      $SO(17)$&6&4.8\%\\
      \hline
      $SO(19)$&8&6.4\%\\
      \hline
      $SO(21)$&6&4.8\%\\
      \hline
      $SO(8)$&8&6.4\%\\
      \hline
      $SO(12)$&35&28\%\\
      \hline
      $SO(14)$&10&8\%\\
      \hline
      $SO(16)$&10&8\%\\
      \hline
      $SO(18)$&14&11.2\%\\
      \hline
      $SO(20)$&12&9.6\%\\
      \hline
      $SO(22)$&9&7.2\%\\
      \hline
      $E_6$&14&11.2\%\\
      \hline
      $E_7$&1&0.8\%\\
      \hline
      $U(1)$&125&100\%\\
      \hline \hline
    \end{tabular}
  \end{subtable}
  \qquad
  \begin{subtable}[t]{7cm}
    \caption{Layer 1, Order 3}
    \label{table:nahe_variations_so10_l1o3_groups}
    \begin{tabular}{||C{2.5cm}|C{2cm}|C{2cm}||}
      \hline \hline
      Gauge Group & Number of Unique Models & \% of Unique Models \\
      \hline \hline
      $SU(2)$&155&57.2\%\\
      \hline
      $SU(3)$&27&9.963\%\\
      \hline
      $SU(4)$&59&21.77\%\\
      \hline
      $SU(5)$&14&5.166\%\\
      \hline
      $SU(6)$&59&21.77\%\\
      \hline
      $SU(7)$&22&8.118\%\\
      \hline
      $SU(8)$&24&8.856\%\\
      \hline
      $SU(9)$&36&13.28\%\\
      \hline
      $SU(10)$&26&9.594\%\\
      \hline
      $SU(11)$&19&7.011\%\\
      \hline
      $SU(12)$&11&4.059\%\\
      \hline
      $SU(13)$&1&0.369\%\\
      \hline
      $SU(14)$&1&0.369\%\\
      \hline
      $SO(8)$&48&17.71\%\\
      \hline
      $SO(12)$&35&12.92\%\\
      \hline
      $SO(14)$&22&8.118\%\\
      \hline
      $SO(16)$&10&3.69\%\\
      \hline
      $SO(18)$&7&2.583\%\\
      \hline
      $SO(20)$&2&0.738\%\\
      \hline
      $SO(22)$&3&1.107\%\\
      \hline
      $E_6$&15&5.535\%\\
      \hline
      $E_7$&4&1.476\%\\
      \hline
      $E_8$&2&0.738\%\\
      \hline
      $U(1)$&271&100\%\\
      \hline \hline
    \end{tabular}
  \end{subtable}
\end{table}

\begin{table}[!ht]
  \renewcommand{\arraystretch}{0.75}
  \caption{The hidden sector gauge group content for the NAHE variation
  extension $L1O3$ models with GUT observable.}
  \label{table:nahe_variations_gut_hidden_sector}
  \begin{subtable}[t]{7cm}
    \caption{Left-Right Symmetric}
    \label{table:nahe_variations_lrsm_hidden_sector}
    \begin{tabular}{||C{2.5cm}|C{2.2cm}|C{2.2cm}||}
      \hline\hline
      Gauge Group & Number of Unique Models & \% of Unique Models \\
      \hline\hline
      $SU(4)$&12&19.67\%\\
      \hline
      $SU(7)$&14&22.95\%\\
      \hline
      $SU(8)$&7&11.48\%\\
      \hline
      $SU(9)$&9&14.75\%\\
      \hline
      $SU(10)$&12&19.67\%\\
      \hline
      $SU(11)$&17&27.87\%\\
      \hline
      $SU(12)$&2&3.279\%\\
      \hline
      $SO(8)$&8&13.11\%\\
      \hline
      $SO(10)$&6&9.836\%\\
      \hline
      $U(1)$&61&100\%\\
      \hline \hline
    \end{tabular}
  \end{subtable}
  \qquad
  \begin{subtable}[t]{7cm}
    \caption{MSSM}
    \label{table:nahe_variations_mssm_hidden_sector}
    \begin{tabular}{||C{2.5cm}|C{2.2cm}|C{2.2cm}||}
      \hline \hline
      Gauge Group & Number of Unique Models & \% of Unique Models \\
      \hline \hline
      $SU(4)$&13&20.63\%\\
      \hline
      $SU(6)$&1&1.587\%\\
      \hline
      $SU(7)$&14&22.22\%\\
      \hline
      $SU(8)$&7&11.11\%\\
      \hline
      $SU(9)$&9&14.29\%\\
      \hline
      $SU(10)$&12&19.05\%\\
      \hline
      $SU(11)$&18&28.57\%\\
      \hline
      $SU(12)$&3&4.762\%\\
      \hline
      $SO(8)$&8&12.7\%\\
      \hline
      $SO(10)$&6&9.524\%\\
      \hline \hline
    \end{tabular}
  \end{subtable}
  \qquad
  \begin{subtable}[t]{7cm}
    \caption{$SU(5) \otimes U(1)$}
    \label{table:nahe_variations_fsu5_hidden_sector}
    \begin{tabular}{||C{2.5cm}|C{2.2cm}|C{2.2cm}||}
      \hline\hline
      Gauge Group & Number of Unique Models & \% of Unique Models \\
      \hline\hline
      $SU(2)$&87&52.73\%\\
      \hline
      $SU(3)$&19&11.52\%\\
      \hline
      $SU(4)$&28&16.97\%\\
      \hline
      $SU(6)$&8&4.848\%\\
      \hline
      $SU(7)$&20&12.12\%\\
      \hline
      $SU(8)$&23&13.94\%\\
      \hline
      $SU(9)$&34&20.61\%\\
      \hline
      $SU(10)$&35&21.21\%\\
      \hline
      $SU(11)$&32&19.39\%\\
      \hline
      $SU(12)$&1&0.6061\%\\
      \hline
      $SO(8)$&22&13.33\%\\
      \hline
      $SO(10)$&14&8.485\%\\
      \hline
      $SO(12)$&7&4.242\%\\
      \hline
      $SO(14)$&5&3.03\%\\
    \hline \hline
    \end{tabular}
  \end{subtable}
  \qquad
  \begin{subtable}[t]{7cm}
    \caption{Pati-Salam}
    \label{table:nahe_variations_ps_hidden_sector}
    \begin{tabular}{||C{2.5cm}|C{2.2cm}|C{2.2cm}||}
      \hline\hline
      Gauge Group & Number of Unique Models & \% of Unique Models \\
      \hline\hline
      $SU(3)$&12&9.6\%\\
      \hline
      $SU(5)$&8&6.4\%\\
      \hline
      $SU(6)$&25&20\%\\
      \hline
      $SU(8)$&29&23.2\%\\
      \hline
      $SU(9)$&24&19.2\%\\
      \hline
      $SU(10)$&15&12\%\\
      \hline
      $SU(11)$&3&2.4\%\\
      \hline
      $SU(12)$&7&5.6\%\\
      \hline
      $SO(8)$&9&7.2\%\\
      \hline
      $SO(10)$&11&8.8\%\\
      \hline
      $SO(12)$&22&17.6\%\\
      \hline
      $SO(14)$&19&15.2\%\\
      \hline
      $SO(16)$&4&3.2\%\\
      \hline
      $SO(20)$&2&1.6\%\\
      \hline
      $E_6$&1&0.8\%\\
      \hline
      $U(1)$&123&98.4\%\\
      \hline \hline
    \end{tabular}
  \end{subtable}
\end{table}

\begin{table}[!ht]
  \centering
    \caption{A near mirrored, NAHE variation order-3 extension.}
    \begin{subtable}[t]{\textwidth}
      \centering
        \caption{Basis vector}
        \label{table:mirrored_model_basis_vectors}
        \begin{tabular}{||c|c||c|c|c|c|c|c|c|c|c|c|c|c|c||}
          \hline\hline
          Sec&O&$\psi$&$x^{12}$&$x^{34}$&$x^{56}$&$\overline{\psi}^{1,...,5}$&$\overline{\eta}^1$&$\overline{\eta}^2$&$\overline{\eta}^3$&$\overline{\phi}^{1,...,8}$&$y^{~12}||\overline{y}^{~12}$&$y^{~34}||\overline{y}^{~34}$&$y^{~56}||\overline{y}^{~56}$&$w^{1,...,6}||\overline{w}^{1,...,6}$\\
          \hline\hline
          $\vec{v}$&3&1&1&0&0&0,...,0&$\frac{2}{3}$&$\frac{2}{3}$&$\frac{2}{3}$&0,...,0,$\frac{2}{3}$,$\frac{2}{3}$,$\frac{2}{3}$&0,0$||$0,0&1,1$||$0,0&1,1$||$0,0&0,0,0,0,0,0$||$0,0,0,0,0,0\\
          \hline
        \end{tabular}
        \vspace{5mm}\\

        $k_{\vec{v},j}$ = (0,0,0,0,0)
    \end{subtable}
    \vspace{5mm}\\
    \begin{subtable}[t]{\textwidth}
      \renewcommand{\arraystretch}{0.65}
      \caption{Particle Content}
      \label{table:mirrored_model_content}
      \begin{tabular}{||c||c|c|c|c|c|c|c|c||}
        \hline \hline
        \textbf{QTY}&$SU(2)$&$SU(2)$&$SU(2)$&$SU(2)$&$SU(2)$&$SU(2)$&$SO(16)$&$SO(16)$\\
        \hline \hline
        1&2&2&2&2&1&1&1&1\\
        \hline
        1&2&2&2&1&1&2&1&1\\
        \hline
        1&2&2&1&2&2&1&1&1\\
        \hline
        1&2&2&1&1&2&2&1&1\\
        \hline
        1&2&2&1&1&1&1&16&1\\
        \hline
        1&2&1&2&2&2&1&1&1\\
        \hline
        1&2&1&2&1&2&2&1&1\\
        \hline
        1&2&1&2&1&1&1&16&1\\
        \hline
        1&2&1&1&2&1&1&1&16\\
        \hline
        1&2&1&1&1&2&1&16&1\\
        \hline
        1&2&1&1&1&1&2&1&16\\
        \hline
        1&1&2&2&2&2&1&1&1\\
        \hline
        1&1&2&2&1&2&2&1&1\\
        \hline
        1&1&2&2&1&1&1&16&1\\
        \hline
        1&1&2&1&2&1&1&1&16\\
        \hline
        1&1&2&1&1&2&1&16&1\\
        \hline
        1&1&2&1&1&1&2&1&16\\
        \hline
        1&1&1&2&2&1&1&1&16\\
        \hline
        1&1&1&2&1&2&1&16&1\\
        \hline
        1&1&1&2&1&1&2&1&16\\
        \hline
        1&1&1&1&2&2&1&1&16\\
        \hline
        1&1&1&1&2&1&2&16&1\\
        \hline
        1&1&1&1&1&2&2&1&16\\
        \hline
        1&1&1&1&1&1&1&128&1\\
        \hline
        1&1&1&1&1&1&1&16&16\\
        \hline
      \end{tabular}
  \end{subtable}
\end{table}

\clearpage

\section{Plots and Figures}

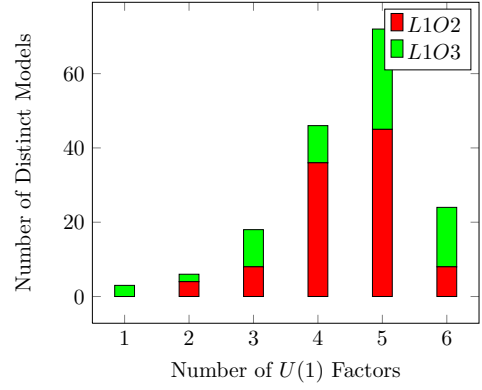
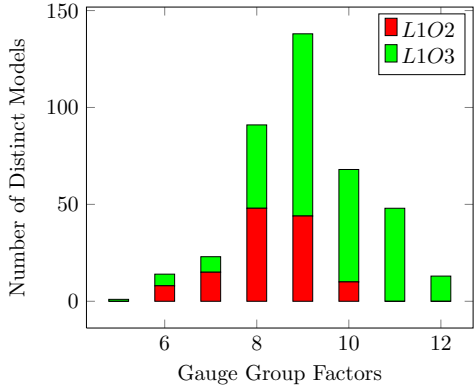
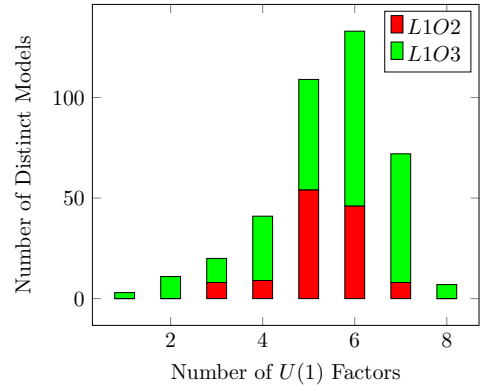
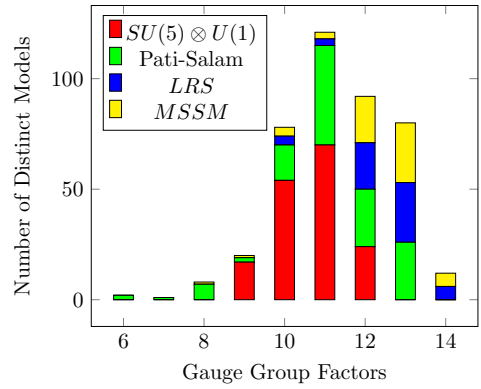
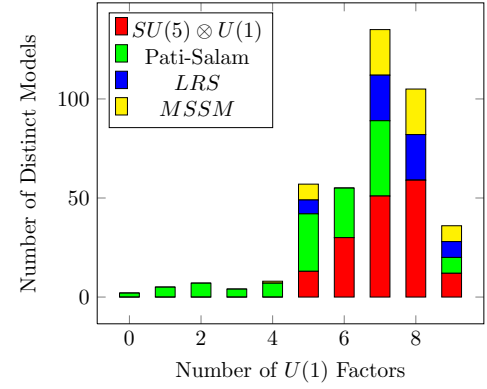
\begin{figure}[!ht]
  \centering
    \begin{subfigure}[b]{0.4\textwidth}
      \begin{tikzpicture}[scale=0.75]
        \begin{axis} [ybar stacked, ylabel = Number of Distinct Models,
                      xlabel = Gauge Group Factors]
          \addplot[draw=black, fill=red]coordinates{
            (5,4) (6,14) (7,38) (8,35) (9,10)(10,0)
          };
          \addlegendentry{$L1O2$}
          \addplot[draw=black, fill=green]coordinates{
            (5,1) (6,3) (7,16) (8,26) (9,16) (10,6)
          };
          \addlegendentry{$L1O3$}
        \end{axis}
      \end{tikzpicture}
      \caption{$E_6$ Gauge Group Factors}
      \label{figure:nahe_variations_e6_group_factors}
    \end{subfigure}
    \hspace{0.1\textwidth}
    \begin{subfigure}[b]{0.4\textwidth}
      \begin{tikzpicture}[scale=0.75]
        \begin{axis} [ybar stacked, ylabel = Number of Distinct Models,
                      xlabel = Number of $U(1)$ Factors]
          \addplot[draw=black, fill=red]coordinates{
            (1,0)(2,4) (3,8) (4,36) (5,45) (6,8)
          };
          \addlegendentry{$L1O2$}
          \addplot[draw=black, fill=green]coordinates{
            (1,3) (2,2) (3,10) (4,10) (5,27) (6,16)
          };
          \addlegendentry{$L1O3$}
        \end{axis}
      \end{tikzpicture}
      \caption{$E_6$ - $U(1)$ Factors}
      \label{figure:nahe_variations_e6_u1_factors}
    \end{subfigure}
    \vspace{5mm}\\
    \begin{subfigure}[b]{0.4\textwidth}
      \begin{tikzpicture}[scale=0.75]
        \begin{axis} [ybar stacked, ylabel = Number of Distinct Models,
                      xlabel = Gauge Group Factors]
          \addplot[draw=black, fill=red]coordinates{
            (5,0) (6,8) (7,15) (8,48) (9,44) (10,10) (11,0) (12,0)
          };
          \addlegendentry{$L1O2$}
          \addplot[draw=black, fill=green]coordinates{
            (5,1) (6,6) (7,8) (8,43) (9,94) (10,58) (11,48) (12,13)
          };
          \addlegendentry{$L1O3$}
        \end{axis}
      \end{tikzpicture}
      \caption{$SO(10)$ Gauge Group Factors}
      \label{figure:nahe_variations_so10_group_factors}
    \end{subfigure}
    \hspace{0.1\textwidth}
    \begin{subfigure}[b]{0.4\textwidth}
      \begin{tikzpicture}[scale=0.75]
        \begin{axis} [ybar stacked, ylabel = Number of Distinct Models,
                      xlabel = Number of $U(1)$ Factors]
          \addplot[draw=black, fill=red]coordinates{
            (1,0)(2,0)(3,8) (4,9) (5,54) (6,46) (7,8)(8,0)
          };
          \addlegendentry{$L1O2$}
          \addplot[draw=black, fill=green]coordinates{
            (1,3) (2,11) (3,12) (4,32) (5,55) (6,87) (7,64) (8,7)
          };
          \addlegendentry{$L1O3$}
        \end{axis}
      \end{tikzpicture}
      \caption{$SO(10)$ - $U(1)$ Factors}
      \label{figure:nahe_variations_so10_u1_factors}
    \end{subfigure}
    \vspace{5mm}\\
    \begin{subfigure}[b]{0.4\textwidth}
      \begin{tikzpicture}[scale=0.75]
        \begin{axis} [ybar stacked, ylabel = Number of Distinct Models,
                      xlabel = Gauge Group Factors, legend pos = north west]
          \addplot[draw=black, fill=red]coordinates{
            (6,0)(7,0)(8,0)(9,17) (10,54) (11,70) (12,24)(13,0) (14,0)
          };
          \addlegendentry{$SU(5)\otimes U(1)$}
          \addplot[draw=black, fill=green]coordinates{
            (6,2) (7,1) (8,7) (9,2) (10,16) (11,45) (12,26) (13,26)(14,0)
          };
          \addlegendentry{Pati-Salam}
          \addplot[draw=black, fill=blue]coordinates{
            (6,0)(7,0)(8,0)(9,0)(10,4) (11,3) (12,21) (13,27) (14,6)
          };
          \addlegendentry{$LRS$}
          \addplot[draw=black, fill=yellow]coordinates{
            (6,0)(7,0)(8,1) (9,1) (10,4) (11,3) (12,21) (13,27) (14,6)
          };
          \addlegendentry{$MSSM$}
        \end{axis}
      \end{tikzpicture}
      \caption{Remaining GUTs -\\ Gauge Group Factors}
      \label{figure:nahe_variations_fsu5_ps_lrs_mssm_group_factors}
    \end{subfigure}
    \hspace{0.1\textwidth}
    \begin{subfigure}[b]{0.4\textwidth}
      \begin{tikzpicture}[scale=0.75]
        \begin{axis} [ybar stacked, ylabel = Number of Distinct Models,
                      xlabel = Number of $U(1)$ Factors,
                      legend pos = north west]
          \addplot[draw=black, fill=red]coordinates{
            (0,0)(1,0)(2,0)(3,0)(4,0)(5,13) (6,30)(7,51)(8,59)(9,12)
          };
          \addlegendentry{$SU(5)\otimes U(1)$}
          \addplot[draw=black, fill=green]coordinates{
            (0,2)(1,5)(2,7)(3,4)(4,7)(5,29)(6,25)(7,38)(8,0)(9,8)
          };
          \addlegendentry{Pati-Salam}
          \addplot[draw=black, fill=blue]coordinates{
            (0,0)(1,0)(2,0)(3,0)(4,0)(5,7)(6,0)(7,23)(8,23)(9,8)
          };
          \addlegendentry{$LRS$}
          \addplot[draw=black, fill=yellow]coordinates{
            (0,0)(1,0)(2,0)(3,0)(4,1) (5,8)(6,0)(7,23)(8,23)(9,8)
          };
          \addlegendentry{$MSSM$}
        \end{axis}
      \end{tikzpicture}
      \caption{Remaining GUTs -\\ U(1) Factors}
    \end{subfigure}
    \caption{Gauge and $U(1)$ statistics for various GUT models in the NAHE variation extensions data set.}
    \label{figure:nahe_variations_gauge_factors}
\end{figure}

\end{document}